\documentclass[twocolumn,prb,showpacs]{revtex4}
\usepackage[pdftex]{graphicx}
\usepackage{dcolumn}
\usepackage{amsmath,amsfonts,amssymb,amsxtra}
\usepackage{mathrsfs}
\usepackage{bm}
\usepackage{times}

\begin{document}

\title{Structural phase transition below 250~K in superconducting
K$_{0.75}$Fe$_{1.75}$Se$_{2}$}

\author{A. Ignatov$^{1}$, A. Kumar$^{1}$, P. Lubik$^{1}$,
R. H. Yuan$^{2}$, W. T. Guo$^{2}$, N. L. Wang$^{2}$, K. Rabe$^{1}$, and G. Blumberg$^{1}$}

\affiliation{$^{1}$Department of Physics and Astronomy, Rutgers, The State University of New Jersey, Piscataway, NJ 08854, USA}
\affiliation{$^{2}$Beijing National Laboratory for Condensed Matter Physics, Institute of Physics, Chinese Academy of Sciences, Beijing 100190, China}

\date{\today}

\begin{abstract}

Vibrational properties of iron-chalcogenide superconductor
K$_{0.75}$Fe$_{1.75}$Se$_{2}$ with $T_{c}\sim$ 30~K have been measured by Raman and optical spectroscopies over temperature range of 3-300~K. Sample undergoes \textit{I4/m}~$\to $ \textit{I4} structural phase transition accompanied by loss of inversion symmetry at $T_{1}$, below 250~K, observed as appearance of new fully-symmetric Raman mode at $\sim$ 165~cm$^{-1}$. Small vibration mode anomalies are also observed at $T_{2}\sim$ 160~K. From first-principles vibrational analysis of antiferromagnetic K$_{0.8}$Fe$_{1.6}$Se$_{2}$ utilizing pseudopotentials all observed Raman and infrared modes have been assigned and the displacement patterns of the new Raman mode identified as involving predominantly the Se atoms.

\end{abstract}

 \maketitle

\section{INTRODUCTION}

Discovery of high-$T_{c}$ superconductivity in iron-based chalcogenides A$_{y}$Fe$_{1.6+x}$Se$_{2}$ (A=K, Rb, Cs, and Tl) Ref. [\onlinecite{discovery}] raised considerable attention since the materials exhibit unusual physical properties. Parent compound ($y=$1, $x=$0) is an insulator, \cite{transport, gap} crystalizes into $\sqrt 5$x$\sqrt 5$x1 \textit{I4/m} Fe vacancy-ordering structure, and exhibits antiferromagnetic (AFM) order below a Ne\'{e}l temperature of $\sim$ 560~K Ref.[\onlinecite{structure}]. Doping with alkaline metals or Tl ($y<$1) apparently preserves the Fe vacancy-ordering and give rise to superconductivity in samples with close to 2:4:5 stoichiometry \cite{compos}. Early transport \cite{transport} and neutron diffraction \cite{structure} studies suggested that the superconductivity coexists with AFM order. Alternatively, the doping is discussed in terms of microscopic phase separation \cite{TEM, XRD, XRD1, NQR0, optics}: a mixture of vacancy-ordered AFM insulating phase and superconducting phase (SC). Due to resent experimental evidences \cite{mSR, NQR, INS, STM} the consensus seems emerge: the AFM and SC phases are specially separated, the AFM phase occupies from $\sim$ 80 Ref. [\onlinecite{mSR}] to 95 \% Ref. [\onlinecite{NQR}] of the sample volume, and SC phase is homogeneous and does not contains Fe-vacancies nor magnetic moments. \cite{NQR, INS, STM}

Raman scattering study of superconducting K$_{0.8}$Fe$_{1.6}$Se$_{2}$ observed at least 13 phonon modes \cite{Raman1}. The crystal symmetry of sample was determined as $C_{4h}$ or lower. Zhang \textit{et al.} performed LDA vibration analysis of nonmagnetic \textit{I4/m} K$_{0.8}$Fe$_{1.6}$Se$_{2}$ phase and assigned majority of observed Raman modes. The vibrational properties K$_{0.88}$Fe$_{1.63}$S$_{2}$ isostructural to K$_{0.8}$Fe$_{1.6}$Se$_{2}$ confirmed Fe-vacancy ordering: 14 Raman active modes predicted by factor-group analysis were observed and assigned. The authors concluded that the phonon energies in the range of 80-300~K are driven by anharmonicity effects without any signatures of electron-phonon interaction \cite{Raman2}. Impact of iron and potassium composition on Raman vibration spectra of A$_{0.8}$Fe$_{1.6}$Se$_{2}$ (A=K, Rb, and Tl) was presented in Ref. [\onlinecite{Raman3}].

The optical studies to date showed at least ten IR-active modes at low temperatures. \cite{optics1,optics2}
The in-plane optical conductivity of ($T_{c}=$31~K) is incoherent at 300~K, dominated by IR-active modes and high-frequency excitations \cite{optics2}, but become coherent just above the $T_{c}$. Small carrier concentration prompted authors \cite{optics, optics2} to suggest that the global superconductivity is due to Josephson coupling of nanoscale-sized superconducting phase in the AFM ordered insulating phase.

In this paper we report on Raman scattering and \textit{ab}-plain optical conductivity studies of superconducting K$_{0.75}$Fe$_{1.75}$Se$_{2}$ ($T_{c}\sim$ 30~ K) in the $T$-range from 3 to 300~K. At least 19 Raman-active and 12 IR-active modes are observed at 3~K. The $\sim$ 136, 143, 242, and 277~cm$^{-1}$ Raman and $\sim$ 208 cm$^{-1}$ IR mode exhibit Fano-like shape. The Raman Fano modes are due the vibration coupling to AFM spin fluctuations, while the IR- mode is coupled to charge carriers in low-frequency part of optical conductivity. Raman phonon linewidth contains approximately equal contributions of two-phonon lattice anharmonicity on one hand and bare self-energy and broadening due to intrinsic defects on the other hand, except for the $\sim$ 100~cm$^{-1}$ mode dominated by inhomogeneous broadening. We show that K$_{0.75}$Fe$_{1.75}$Se$_{2}$ undergoes \textit{I4/m }~$\to $ \textit{I4}  structural phase transition at $T_{1}$ below 250~K. Several modes which are not Raman- and IR-active in the measured geometry in \textit{I4/m} become clearly visible in \textit{I4} phase. Symmetry of the Se-Fe slab is broken at $T_{1}$. At $T_{2}\sim$ 160~K Raman vibration modes exhibit weak anomalies seen as small discontinuity of vibration frequencies and change in vibration intensity vs temperature dependencies. Raman vibration intensities of a few modes increases between $T_{1}$ and $T_{2}$, saturating above the $T_{2}$, except for three modes dominated by $c$-axis atomic displacements: $c$-axis structural distortions within the slab appear to build up on cooling down to 3~K. The low-frequency optical conductivity displays weak temperature dependence above $T_{1}$ followed by faster increase below the $T_{2}$.

\section{EXPERIMENTAL}

The crystal of iron-chalcogenide superconductors were grown by a self-melting method with nominal concentration of 0.8:2.1:2.0 (K:Fe:Se). The actual chemical composition was determined by EDXS as K$_{0.75}$Fe$_{1.75}$Se$_{2}$ (KFS). Two-step transitions were seen in resistivity curve \cite{optics}, a sharp drop at 42 K is followed by a major superconducting transition with $T_{c}\sim$ 30~K. Further details of sample characterization can be found elsewhere. \cite{optics}

KFS crystals were never exposed to air. Sealed vial was open under 99.999{\%} N$_{2}$, crystal removed and glued on replaceable copper sample holder of a helium Oxford Instruments cryostat, dried and cleaved along the \textit{ab}-plane, transferred to the He-flow cryostat, and quickly cooled below water freezing temperature. Raman data were obtained on two single crystals. Data presented in this paper refer to the sample with more detail temperature dependence records. It is worth mentioning that results obtained on the second sample are consistent with findings reported here.

Raman spectra were excited with Kr$^{+}$ laser line of $\lambda =$647.1~nm ($E=$1.92~eV) with less than 10~mW of incident laser power focused into a spot of $\sim$ 50x100~$\mu $m$^{2}$ on the freshly cleaved \textit{ab}-plain crystal surface. The scattered light collected close to the backscattered geometry was focused to a 100x240 $\mu $m$^{2}$ entrance slits of a custom triple-stage spectrometer equipped with 1800~lines/mm gratings. The instrumental resolution was $\sim$ 1.4~cm$^{-1}$. To record symmetry resolved Raman spectra we employed circularly polarized light with the optical configurations selecting either the same or opposite chirality for incident and scattered light. The former is referred to as right-right (RR) and the latter as to right-left (RL) configurations. For the $C_{4h}$ point group the $B_{g}$ ($A_{g})$ symmetry is probed in the RL (RR) scattering geometry. Temperature dependent Raman spectra were collected at 3, 45, 100, 150, 160, 180, 200, 260, and 300~K with $T$- stability better than 0.1~K. An estimated local heating in the laser spot did not exceed 4~K.

Optical measurements were done by Bruker Vertex 80v spectrometer in the frequency range from 25 to 10000~cm$^{-1}$. The sample was under vacuum of 2*10$^{-5}$ ~Pa. An in-situ gold and aluminum over-coating technique was used to get the reflectance $R(\omega )$ for light polarized in the KFS (\textit{ab})- planes. The real part of conductivity $\sigma_{1}(\omega )$ was obtained by the Kramers-Kronig transformation of the $R(\omega )$. Optical spectra were collected at 8, 35, 170, and 300~K.

\section{RESULTS}
\begin{figure*}
\includegraphics[width=18.0cm]{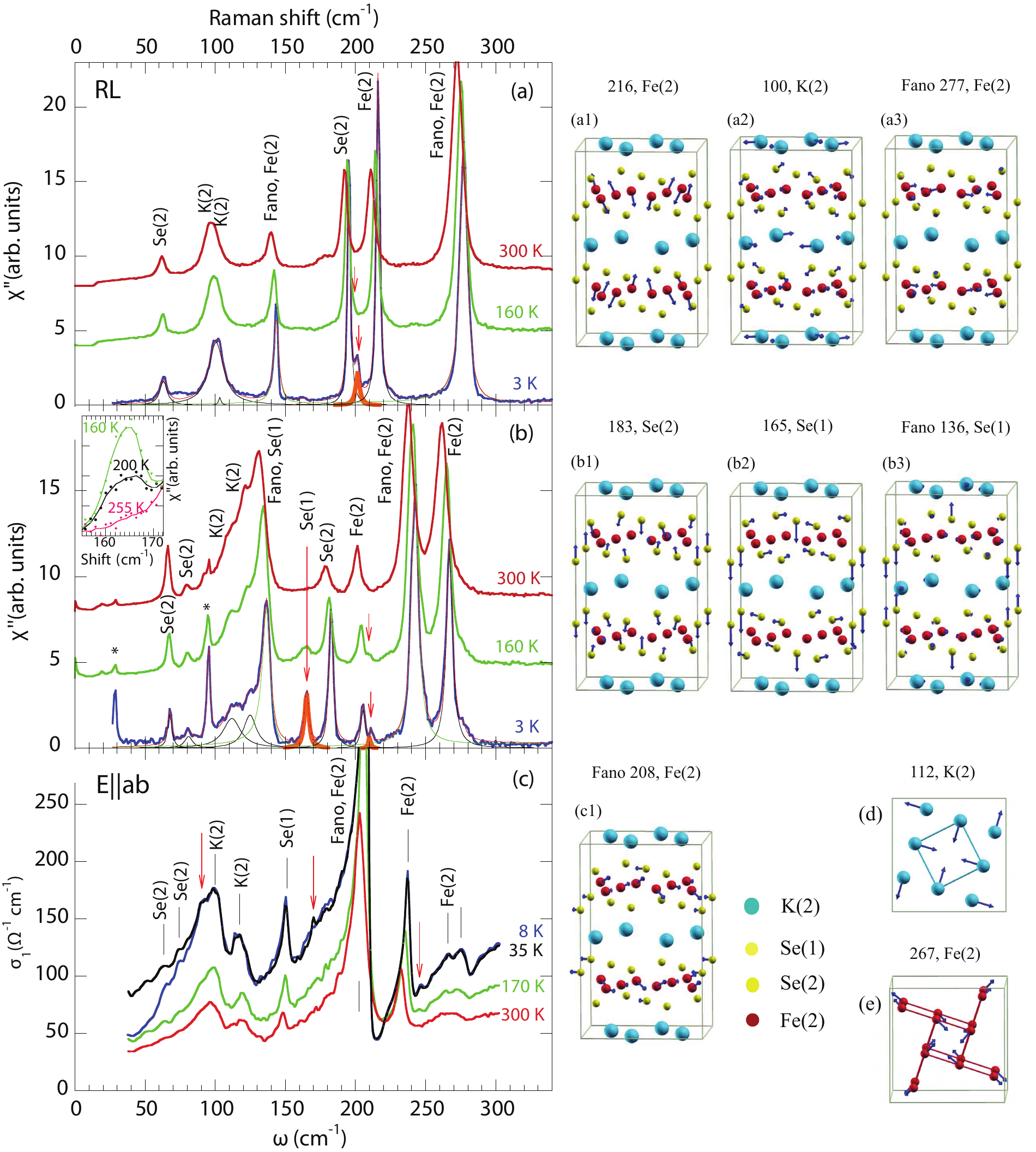}
\caption{(Color online) Raman response function of
K$_{0.75}$Fe$_{1.75}$Se$_{2}$ in (a) RL and (b) RR channels at 3 (blue), 160 (green), and 300~K(red). (c) \textit{ab}-plane optical conductivity of K$_{0.75}$Fe$_{1.75}$Se$_{2}$ from the same batch of samples at 8, 35, 170, and 300~K, adopted from Fig. 2 in Ref. [\onlinecite{optics}]. Fits are illustrated on the 3 K spectra. Modes belonging to \textit{I4/m} Ref. [\onlinecite{structure}] are shown in black (Lorentz) and green (Fano). New phonons appearing at $T_{1}$ in the range of 200 to 255~K [insert in (b)], usually referred as seen below $\sim$250~K, are shown in orange and marked with arrows. Predominant atomic contribution is labeled for each vibration mode.
Computed displacement patterns of selected Raman modes in AFM \textit{I4/m}, $\sqrt 5$x$\sqrt 5$ unit cells of K$_{0.8}$Fe$_{1.6}$Se$_{2}$ are shown to the right from the vibration spectra: The $B_{g}$ modes at $\sim$ 216 and 100~cm$^{-1}$ and Fano mode at 277~cm$^{-1}$ are visualized in (a1)-(a3). The displacement patterns of $A_{g}$ mode at $\sim$ 183,  $A_{u}$ mode at $\sim$ 165 (becomes Raman active $A$ mode below $T_{1}$), and Fano $A_{g}$ mode mode at $\sim$ 136 cm$^{-1}$ are shown in (b1), (b2), and (b3). Nine IR-active modes observed throughout the $T$-range of this study are marked with black bars in (c). At least three new modes (red arrows) seen at 170~K. The displacement patterns for IR Fano mode at $\sim$ 208~cm$^{-1}$ is shown in (c1). 2D atomic displacement patters of $A_{g}$ modes involving both chiral and breathing displacements: (d) K(2)-based $\sim$ 112~cm$^{-1}$ and (e) Fe(2) 267 cm$^{-1}$. All displacements are not up to scale. Two peaks at $\sim$ 29 and 95~cm$^{-1}$ marked by $^{*}$ in (b) are due to plasma lines of Kr- laser.}
\label{fig1}
\end{figure*}

K$_{0.8}$Fe$_{1.6}$Se$_{2}$ crystalizes into tetragonal structure \textit{I4/m} (space group {\#}87) Ref. [\onlinecite{structure}], resulting in the irreducible vibrational representation:
\begin{equation}
\Gamma_{\text{vib}} = 9A_{g} \oplus 8B_{g} \oplus 8E_{g} \oplus 9A_{u}  \oplus 7B_{u} \oplus 10E_{u}
\end{equation}
All $g$-modes are Raman active, but only the $A_{g}$ and $B_{g}$ are selected with RR and RL polarizations under the measured geometry. The $A_{u}$ and $E_{u}$ vibrations are infrared active along the $c$-axis and in the $ab$-plane. The $B_{u}$ modes are silent. The Fe(1)-vacancy related vibration modes are excluded.
Throughout this paper we adopted commonly used site designation: K(1), K(2), Fe(1), Fe(2), Se(1) and Se(2)
stand, respectively, for Wyckoff positions of 2$a$, 8$h$, 4$d$, 16$i$, 4$e$, and 16$i$, refer to a legend of Fig.\ref{fig1}.

Raman spectra of
K$_{0.75}$Fe$_{1.75}$Se$_{2}$ are shown in Fig~\ref{fig1}(a) and \ref{fig1}(b) for RL and RR polarizations, respectively. At 300~K, at least 7(9) modes are observed in RL (RR) in good agreement with 9$A_{g}$+8$B_{g}$
expected in the \textit{I4/m} $\sqrt 5$x$\sqrt 5$ cell of K$_{0.8}$Fe$_{1.6}$Se$_{2}$ \cite{Raman1}.
The $A_{g}$ modes at $\sim$ 112 and 267~cm$^{-1}$ are dominated by both chiral and breathing displacements of K(2) and Fe(2) atoms, respectively, Fig.~\ref{fig1}(d,e). Below 200~K new modes appear (marked with red arrows): at $\sim$ 201~cm$^{-1}$ in RL and at $\sim$ 165 and $\sim$ 211~cm$^{-1}$ in the RR. The $\sim$ 136 and $\sim$ 277~cm$^{-1}$ phonons in the RL and $\sim$ 144 and $\sim$ 242~cm$^{-1}$ phonons in RR exhibit Fano shapes in whole temperature range of this study. The Fano modes become more symmetric with temperature decrease. Phonon modes parameters are derived from least square fit to experimental data and are summarized in Table \ref{tab1}.

Low-frequency region of optical conductivity adopted from Fig. 2 in Ref. [\onlinecite{optics}] is shown in Fig~\ref{fig1}(c). In agreement with previous studies \cite{optics, optics2}, $\sigma_{1}(\omega )$ is small (characteristic of a poor metal) and it is dominated by the infrared-active vibrations and interband features at higher energies. At 170 and 300~K 9 IR-active modes are observed. The $\sim$ 208~cm$^{-1}$ mode exhibits a Fano-like shape, becoming more asymmetric on cooling. At 170~K and below, at least three new modes (red arrows) are formed, Fig.~\ref{fig1}(c). An inspection of the Table \ref{tab1} reveled that Raman and IR modes reported in this work do not overlap. It's therefore tempting to conclude that inversion symmetry is preserved. In Section IV we argue that inversion symmetry is actually broken below $T_{1}\sim$ 250~K. The conductivity displays relatively weak temperature dependence above 170~K followed by about two-fold increase of the continuum as temperature drops from 170 to 35~K. In agreement with previous studies \cite{optics1, optics2}, a Drude-like peak is seen in 35~K data, shortly before sample becomes superconducting.

\begin{table}
\begin{center}
\caption {Assignment of observed Raman and IR vibration modes in K$_{0.75}$Fe$_{1.75}$Se$_{2}$ based on
comparison with first-principle calculations utilizing pseudopotentials. Raman and IR modes which appear below $T_{1}\sim$ 250~K are marked by $^{+}$ and $^{\#}$. The Lorentz parameters for Raman and \textit{ab}-plane IR modes, respectively, at 3 and 35~K were obtained from fit to the experimental data shown in Fig.~\ref{fig1}. Here $\omega_{i}$ and $\Gamma_{i}$(in cm$^{-1}$) are the frequency and FWHM, of the $i$-th mode.
Error bars estimated from covariance are 0.2-0.4 and 0.4-2.0 cm$^{-1}$ for $\omega_{i}$ and $\Gamma_{i}$.
Raman data from Ref. [\onlinecite{Raman1}] and IR data from Ref. [\onlinecite{optics2}] are shown for comparison.
Computed vibration frequencies are shown for \textit{I4/m }structure.
Total number of modes is less that in Eq(1) because K(1) and acoustic modes are not computed/omitted.
Vibration frequencies in \textit{I4} are less than 10~cm$^{-1}$ apart are not shown in this Table.
$A_{g}$, $B_{g}$, and $E_{u}$ modes in \textit{I4/m } becomes, respectively $A$, $B$, and $E$ modes in \textit{I4},
refer to Sect IV.}
\begin{tabular}{p{55pt} c c c c }
\hline \hline
AFM, \textit{I4/m}&
Raman~($\omega $, $\Gamma )$&
Ref.[\onlinecite{Raman1}]&
IR~($\omega $, $\Gamma )$&
Ref.[\onlinecite{optics2}] \\
\hline
63.6 $A_{g}$&
67.6, 3.5&
66.3&
&
 \\
%\hline
79.9 $A_{g}$&
81.0, 6.1&
&
&
 \\
%\hline
89.0 $A_{g}$&
111.8, 15.0&
&
&
 \\
%\hline
108.2 $A_{g}$&
124.8, 11.1&
123.8&
&
 \\
%\hline
126.0 $A_{g}$&
Fano 135.9&
134.6&
&
 \\
%\hline
173.5 $A_{g}$&
182.5, 3.4&
&
&
 \\
%\hline
211.1 $A_{g}$&
205.3, 3.6&
202.9&
&
 \\
%\hline
236.0 $A_{g}$&
Fano 242.3&
239.4&
&
 \\
%\hline
265.9 $A_{g}$&
267.0, 5.3&
264.6&
&
 \\
\hline
57.9 $B_{g}$&
63.1, 6.2&
61.4&
&
 \\
%\hline
66.4 $B_{g}$&
- -&
&
&
 \\
%\hline
98.2 $B_{g}$&
100.6, 12.8&
100.6&
&
 \\
%\hline
117.3 $B_{g}$&
103.3, 2.1&
&
&
 \\
%\hline
134.9 $B_{g}$&
Fano 143.6&
141.7&
&
 \\
%\hline
206.0 $B_{g}$&
195.3, 2.6&
&
&
 \\
%\hline
224.0 $B_{g}$&
216.1, 3.0&
214.3&
&
 \\
%\hline
262.7 $B_{g}$&
Fano 277.1&
274.9&
&
 \\
\hline
59.0 $E_{g}$&
&
&
&
 \\
%\hline
79.9 $E_{g}$&
&
&
&
 \\
%\hline
95.1 $E_{g}^{\#}$&
&
&
98.9, 8.2&
102.3 \\
%\hline
104.5 $E_{g}$&
&
&
&
 \\
%\hline
156.9 $E_{g}^{\#}$&
&
&
171.2, 4.9&
 \\
%\hline
206.5 $E_{g}$&
&
&
&
 \\
%\hline
224.0 $E_{g}$&
&
&
&
 \\
%\hline
251.3 $E_{g}^{\#}$&
&
&
246.3, 5.&
 \\
\hline
61.1 $A_{u}$&
&
&
&
 \\
%\hline
92.7 $A_{u}$&
&
&
&
 \\
%\hline
96.3 $A_{u}$&
&
&
&
 \\
%\hline
172.4 $A_{u}^{+}$&
165.2, 3.6&
&
&
 \\
%\hline
212.4 $A_{u}^{+}$&
211.0, 1.4&
&
&
 \\
%\hline
249.7 $A_{u}$&
&
&
&
 \\
%\hline
271.0 $A_{u}$&
&
&
&
 \\
\hline
67.7 $B_{u}$&
&
&
&
 \\
%\hline
76.0 $B_{u}$&
&
&
&
 \\
%\hline
89.3 $B_{u}$&
&
&
&
 \\
%\hline
116.2 $B_{u}$&
&
&
&
 \\
%\hline
181.9 $B_{u}^{+}$&
200.5, 3.4&
&
&
 \\
%\hline
240.5 $B_{u}$&
&
&
&
 \\
%\hline
273.0 $B_{u}$&
&
&
&
 \\
\hline
62.5 $E_{u}$&
&
&
63.8, 7.1&
65.2 \\
%\hline
73.9 $E_{u}$&
&
&
74.2, 6.1&
73.6 \\
%\hline
85.9 $E_{u}$&
&
&
91.3, 16.&
93.7 \\
%\hline
96.5 $E_{u}$&
&
&
118.5,10.&
121.9 \\
%\hline
140.7 $E_{u}$&
&
&
150.1, 4.0&
151.7 \\
%\hline
219.3 $E_{u}$&
&
&
Fano 207.6&
208.3 \\
%\hline
233.3 $E_{u}$&
&
&
236.6, 4.4&
238.3 \\
%\hline
268.3 $E_{u}$&
&
&
267.2, 6.1&
267.1 \\
%\hline
&
&
&
276.1, 4.9&
278.6 \\
\hline \hline
\end{tabular}
\label{tab1}
\end{center}
\end{table}

\subsection{First-principles phonon modes analysis}

We performed first-principles density functional theory (DFT) calculations
using local density approximation (LDA) with Perdew Zunger (PZ) parameterization
for the exchange-correlation energy functional as implemented in \textit{Quantum Espresso}
simulation package \cite{PWSCF}. We used ultrasoft pseudopotentials \cite{USP} for K, Fe
and norm-conserving pseudopotential \cite{NCP} for Se to describe the interaction between
the ionic cores and the valence electrons. The pseudopotentials include 9 valence
electrons for K ($3s^{2}$, $3p^6$, $4s^2$), 16 for Fe ($3s^{2}$, $3p^6$, $3d^6$, $4s^2$),
and 6 for Se ($4s^{2}$, $4p^4$) atoms. We used a plane wave basis with energy
cutoff of 40~Ry for wave function and 360~Ry for the change density and a
$6\times 6\times 4$ Monkhorst Pack \cite{mh-pack} $k$-point mesh for the Brillouin zone (BZ) integration.

We optimized structure of K$_{0.8}$Fe$_{1.6}$Se$_{2}$ with a four spin cluster AFM
ordering as discussed by Bao \textit{et al}. \cite{structure} using experimental lattice constants
obtained at 11~K. The calculation was done using a primitive unit cell
of 22 atoms with K(1) and Fe(1) vacancies at 2$a$ and 4$b$ Wyckoff sites of
space group \textit{I4/m} respectively. Structural optimization is carried through minimization of
energy using Hellman-Feynman forces at each atoms in Broyden-Flecher-Goldfarb-Shanno scheme.
The optimized structure shows a very good agreement with the experiment. We
also allowed inversion symmetry breaking displacement and found that \textit{I4} structure
has slightly lower energy (~3.5~meV) compared to \textit{I4/m}, however the splitting of atomic
coordinates was very small. We find that both the structures exhibit a band gap of $\sim$ 0.4~eV.

Frequencies of the zone center phonons are determined using linear response
method \cite{LR} for the relaxed structures are listed in the first column of Table
\ref{tab1} for the AFM \textit{I4/m} K$_{0.8}$Fe$_{1.6}$Se$_{2}$. Since all nine $A_{g}$ modes
anticipated in the parent AFM \textit{I4/m} K$_{0.8}$Fe$_{1.6}$Se$_{2}$ are observed at room
temperature, there is a unique correspondence between computed and measured vibration
frequencies summarized in Table \ref{tab1}. Computed $B_{g}$ mode at 66.4~cm$^{-1}$ is not observed.
The $B_{g}$ vibration at $\sim$ 117.3 cm$^{-1}$ submerges to the broad $B_{g}$ mode at $\sim$ 100~cm$^{-1}$, Fig. \ref{fig1}(a)
and is revealed in the fit. Importantly, two $A_{g}$ modes at 79.9 and 89.0~cm$^{-1}$cannot be reproduced
in non-magnetic (NM) \textit{I4/m} $\sqrt 5$x$\sqrt 5$ structure either undoped K$_{0.8}$Fe$_{1.6}$Se$_{2}$ or vacancy-free
KFe$_{2}$Se$_{2}$ also computed in this work but not listed in the Table \ref{tab1}. Therefore, accounting for the
spin degree of freedom is essential for accurate mapping of the observed Raman modes. The 8 out of 9
observed IR active modes at 300~K are assigned, Table \ref{tab1} and Fig. \ref{fig1}(c). The remaining $E_{u}$ mode at $\sim$ 278~
cm$^{-1}$ is likely due to finite Fe(1) population in the superconducting sample: an extra $E_{u}$
mode appears in NM vacancy-free \textit{I4/m} at 294~cm$^{-1}$. In summary, the observed Raman and IR-
vibration frequencies above $\sim$ 200 - 250~K are in good agreement with computed frequencies. Below 200 - 250~K, new
Raman modes at $\sim$ 165, 201, and 211~cm$^{-1}$ and IR-active modes at $\sim$ 99, 171, and 246~cm$^{-1}$
show up. Their vibration frequencies corresponds well to the computed frequencies of Raman active A,
B and IR active E in the \textit{I4} structure, Table \ref{tab1}.

Displacement patterns of selected vibration modes are illustrated in Fig. \ref{fig1}.
Raman modes shown in (a1) and (b1) correspond to Fe B$_{1g}$ and As A$_{1g}$ vibrations in the 122 iron-arsenides \cite{modes122}.
In AFM \textit{I4/m} both modes got finite ($x$,$y$) displacements.
The B$_{g}\sim$ 100 cm$^{-1}$ mode is dominated by in-plane K displacements with
some admixture of Se and Fe displacements. Its large $T$-independent linewidth is related to
static disorder associated with K(2) sites. The $A_{u}\sim$ 172.4~cm$^{-1}$ patterns are shown in
(b2). Being non-Raman active in the AFM \textit{I4/m } phase it becomes new Raman mode $A$ in the low-$T$ phase
AFM \textit{I4}, Fig. \ref{fig1}(b2). Fig. \ref{fig1} visualizes atomic displacement of one
IR-active (c1) and two out of four observed Raman active Fano modes (a3 and b3) discussed in this paper.
The striking feature of all but Se-based $A_{g}\sim$ 136~cm$^{-1}$ is essential involvement of Fe(2) atomic displacements.
In K$_{0.8}$Fe$_{1.6}$Se$_{2}$ Fe atoms carry magnetic moment as large as 3.3$\mu_{B}$ Ref. [\onlinecite{structure}],
while electronic structure near $E_{f}$ is dominated by Fe $d$-states \cite{gap}.
The Fano modes coupling to electronic and magnetic degrees of freedom are explored in the following section.

\subsection{Origin of Fano vibration modes}

\begin{figure}
\includegraphics[width=8.5cm]{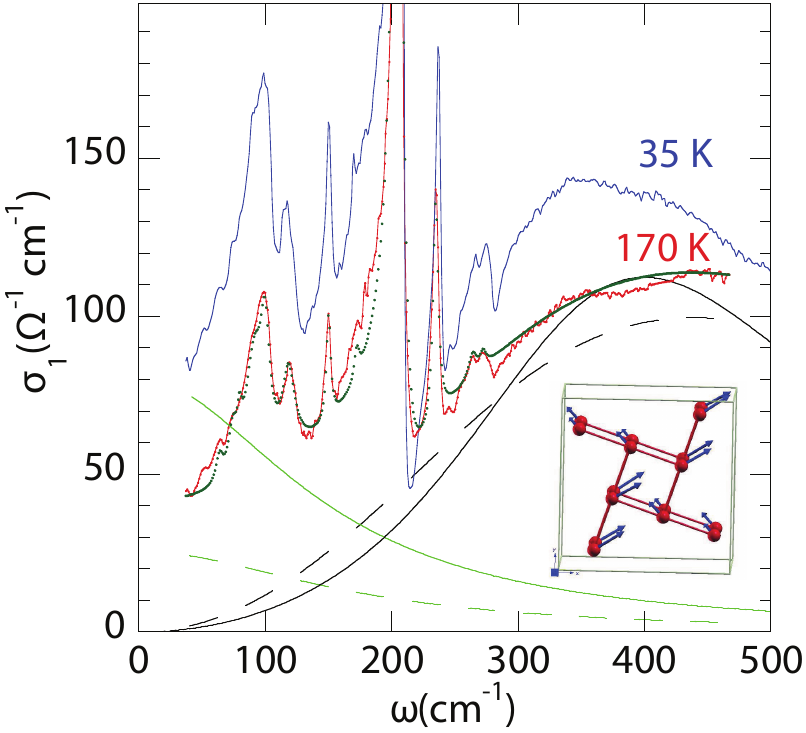}
\caption{(Color online) Optical conductivity at 35~K (blue) and 170~K (red) along with a fitting curve (dark green dots). Drude components (green) and MIR Lorentz terms (black) are shown with solid(dashed) curves for 35(170)~K. Insert: c-axis view of \textit{ab}- plane displacement patterns (not up to scale) of the Fe(2) atoms.}
\label{fig2-ir}
\end{figure}

\begin{figure}
\includegraphics[width=8.5cm]{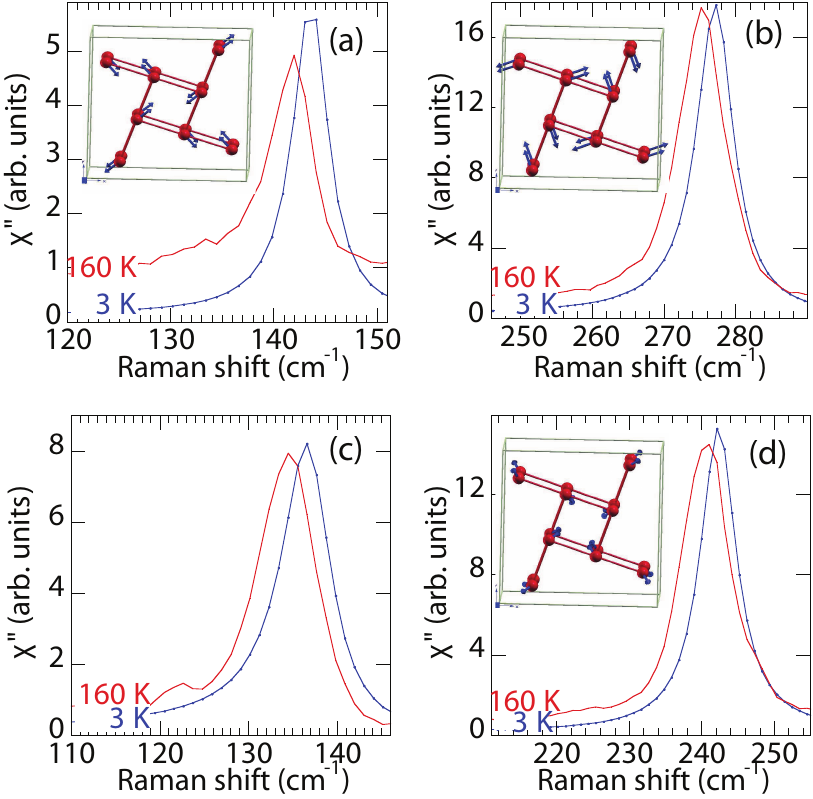}
\caption{(Color online)
Raman Fano modes for (a) $\sim$ 142 and (b) $\sim$ 277~cm$^{-1}$ in RL and (c) $\sim$ 137 and (d) $\sim$ 242~cm$^{-1}$ in RR. Inserts: $c$-axis view of \textit{ab}- plane displacement patterns (not up to scale) of the Fe(2) atoms. The 3D displacement patterns are shown in Fig. \ref{fig1}.}
\label{fig2}
\end{figure}

Asymmetric line shapes are characteristic of Fano resonances arising from coupling between the phonons and an electronic continuum, electronic or magnetic in origin. Dipole transition in the IR- absorption does not directly couple the AFM excitations, but couples charge carriers. Raman scattering probes both electronic and magnetic excitations. The 208~cm$^{-1}$ IR- and $\sim$ 144, 242, and 277 cm$^{-1}$ Raman Fano modes are dominated by Fe(2) atomic displacements. Interestingly, the IR mode gets more asymmetric [Fig. \ref{fig2-ir}] while four Raman modes [Fig. \ref{fig2}(a-d)] become more symmetric on cooling.

The optical conductivity at 35 and 170~K is shown in Fig. \ref{fig2-ir}. Taking 170 K spectrum as an example, the experimental conductivity is fitted as sum of Drude peak (dashed green), broad Lorentz component (dashed black) describing interband transition at $\sim$ 400~cm$^{-1}$, the beginning of mid-infrared (MIR) peak, and eleven Lorentz and one Fano phonon modes. Both Drude and MIR become slightly more coherent and better pronounced as temperature decreases to 35~K. At frequency of Fano mode, the Dude contribution increases, while the MIR contribution slightly decreases. Therefore, enhancement in the IR Fano peak asymmetry is due to the vibration coupling to charge carriers in the Drude tail.

Four Fano Raman modes are presented in Fig. \ref{fig2}(a)-(d) at 3 and 160~K. They were obtained by removal of fitted in phonons from data shown in Fig 1(a,b). Clearly, all Raman modes exhibit similar $T$ dependence: (a) they are less symmetric, and (b) they characterized by larger background at 160~K than at 3~K. The observed behavior is reminiscent of $T$-dependence of a new mode observed in AFM 122 systems. The mode appears at $T_{N}$ as a Fano-shaped one and it becomes progressively more symmetric with temperature decrease. The Fano peak derives from vibration coupling to magnetic continuum, the AFM spin fluctuations.

\subsection{Temperature dependence of Raman mode linewidth and phonon frequencies}

\begin{figure}
\includegraphics[width=8.5cm]{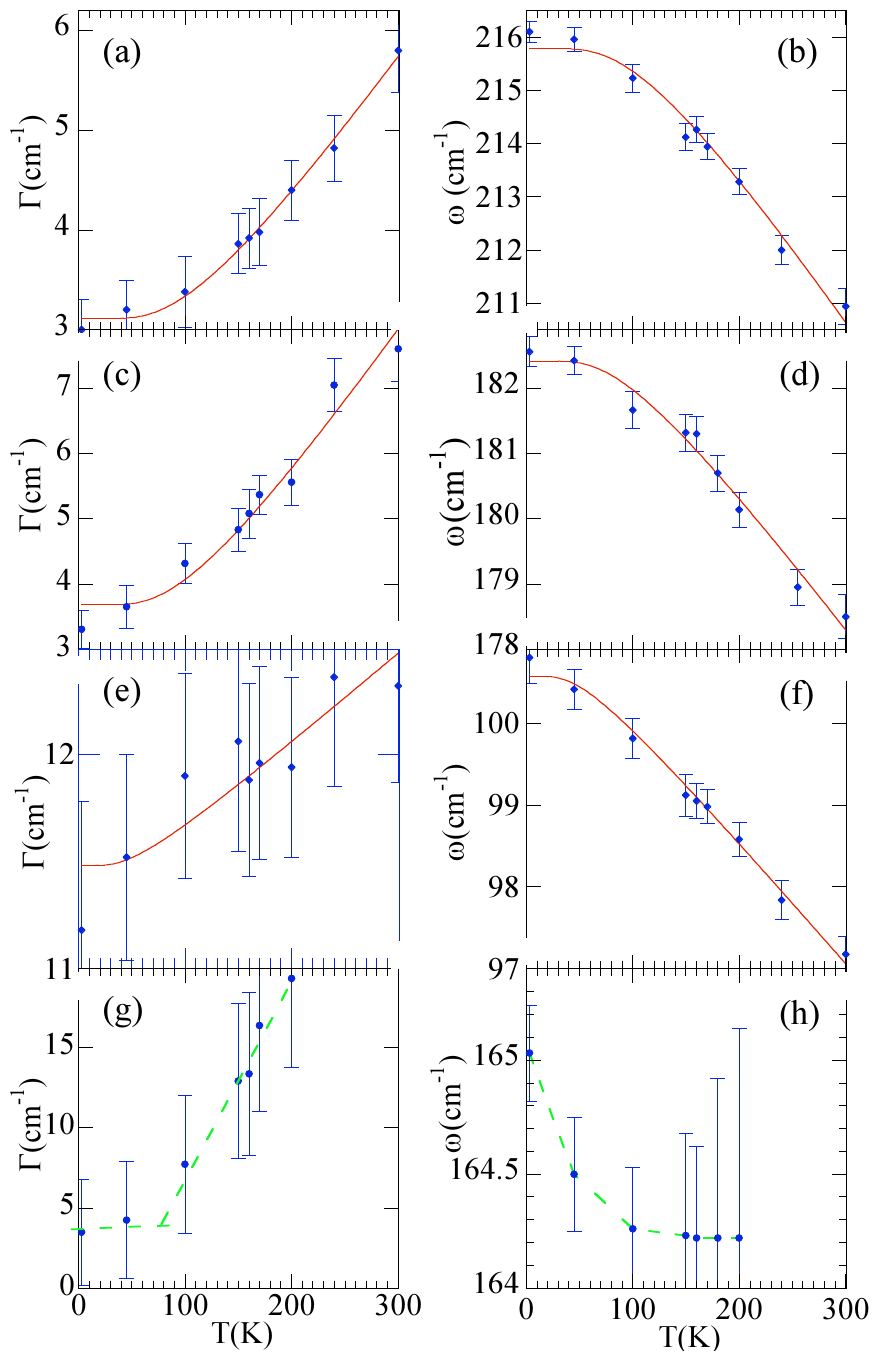}
\caption{(Color online) Linewidth (first row) and subsequent phonon
frequencies (second row) of Fe(2) $B_{g}\sim$ 216, Se(2) $A_{g}\sim$ 182, K(2) $B_{g}\sim$ 100, and Se(1) $A_{g}\sim$ 165~cm$^{-1}$ modes. Solid red lines describe two-phonon anharmonic decay \cite{gamma}. Dashed green lines are guided to the eye. The atomic displacement patterns of the modes are visualized in Fig ~\ref{fig1} (a1, b1, a2, and b2).}
\label{fig3}
\end{figure}

Selected linewidth and phonon frequencies as function of temperature are shown in Fig. \ref{fig3}. The $T$-dependent phonon frequencies qualitatively agree with those reported by Zhang \textit{et al} in
K$_{0.8}$Fe$_{1.6}$Se$_{2}$ (Fig 5 in Ref. [\onlinecite{Raman1}]) and by Lazarevi\'{c} \textit{et al} in isostructural K$_{0.88}$Fe$_{1.63}$S$_{2}$ (Fig. 3(b-j) in Ref. [\onlinecite{Raman2}]). In the latter work, the authors concluded that the Raman active phonon energies in the range of 80-300~K are fully driven by anharmonicity effects \cite{Raman2}. Interpretation offered in the present work is different: the residual linewidth is compatible [Fig. \ref{fig3}(a,c)] or larger [Fig. \ref{fig3}(e)] than the temperature dependent increment between 3 and 300~K. Therefore, self-energy of non-Fano phonons (i.e. at $\sim$ 195 and 216 cm$^{-1}$ in $B_{g}$ and at $\sim$ 68, 205, and 267~cm$^{-1}$ in $A_{g}$ channels) consist of approximately equal contributions of two-phonon lattice anharmonicity on one hand and bare self-energy and broadening due to intrinsic defects on the other. Self-energy of $\sim$ 100~cm$^{-1}$ mode involving the K(2) atomic displacements is dominated by inhomogeneous broadening. The new $\sim$ 165~cm$^{-1}$ mode appearing at $T_{1}$ in the range of 200 to 250~K becomes fully coherent below $\sim$ 40-60~K: the linewidth presented in Fig. \ref{fig3}(g) quickly reduces by $\sim$ 5 times as temperature decreases from 200 to 60~K, followed by saturation below $\sim$ 40~K. The mode hardens on cooling by $\sim$ 1.0~cm$^{-1}$ [Fig. \ref{fig3}(h)] in the traceable $T$-range.

\section{DISCUSSION}

The new 165~cm$^{-1}$ mode appearing at $T_{1}$ in the range of 200-250~K, [insert of Fig. \ref{fig1}(b)], usually referred as seen at $T_{1}$ below $\sim$ 250~K throughout this paper, is not Raman active in the \textit{I4/m} phase (Table \ref{tab1}), fully symmetric in character, and it quickly becomes coherent with $T$-decrease [Fig. \ref{fig2}(g)]. The question arises whether this mode signifies crystal symmetry lowering on a structural phase transition. If it is associated with symmetry lowering, it would become allowed phonon mode in one of subgroups of $C_{4h}$. The $C_{4h}$ encompass the $C_{4}$(loss of inversion and rotation-reflection), $C_{2h}$(loss of 4$^{th}$ order rotation axis), $S_{4}$ (loss of inversion and 4$^{th}$ order rotation axis),  C$_{2}$ (loss of inversion, 4$^{th}$ order rotation axis, and rotation-reflection), and $C_{1}$(primitive) subgroups.
We did not observe leaking of four-fold axis symmetry that would results in the cross-polarization intensity leakages beyond small leakages of polarization optics which don't correlate with temperature dependence of the 165~cm$^{-1}$ mode. Thus, $C_{2h}$, $C_{2}$, and $C_{1}$ subgroups are excluded. The $S_{4}$ (space group {\#}82) is excluded because there is no new $A$-type Se(1) mode associated with the transition. Therefore,  \textit{I4/m }($C_{4h}$, space group {\#}87) becomes \textit{I4 }($C_{4}$, space group {\#}79). From the factor-group analysis $A_{g}$, $B_{g}$, and $E_{u}$ modes in $C_{4h}$ becomes, respectively $A$, $B$, and $E$ modes in $C_{4}$. Instead of 9$A_{g}+$8$B_{g}$ Raman active and 8$E_{u}$ infrared active modes in high-$T$ \textit{I4m} phase one would expect to encounter 17$A$+15$B$ Raman and 17$E$ IR-active modes in low-$T$ phase under the measured geometry. Here we excluded acoustic and Fe(1)-related modes, since the Fe(1)-site is mostly empty site in K$_{0.75}$Fe$_{1.75}$Se$_{2}$. The Raman active modes do not overlap with \textit{ab}-plane IR active modes, not only in high- but in low-$T$ phases. This explains seemingly puzzling absence of IR mode leakages into the Raman spectra noted in Section 3. Since new Raman modes (at $\sim$ 165, 201, and 211~cm$^{-1}$ and IR-active modes (at $\sim$ 99, 171, and 246~cm$^{-1})$ appears below $T_{1}$ and those modes are non-Raman(non-IR) active $A_{u}(E_{g})$ or silent $B_{u}$ in \textit{I4m} (Table \ref{tab1}) we suggest that K$_{0.75}$Fe$_{1.75}$Se$_{2}$ undergoes \textit{I4/m} $\to $ \textit{I4} structural phase transition accompanied by loss of inversion symmetry at $T_{1}$ below $\sim$ 250~K. Our first-principles calculations utilizing pseudopotentials also narrowly favor \textit{I4} over the \textit{I4/m} structure. The small total energy difference is likely because computations do not include all correlations and/or the calculations are performed for the undoped K$_{0.8}$Fe$_{1.6}$Se$_{2}$.

\begin{figure}
\includegraphics[width=8.5cm]{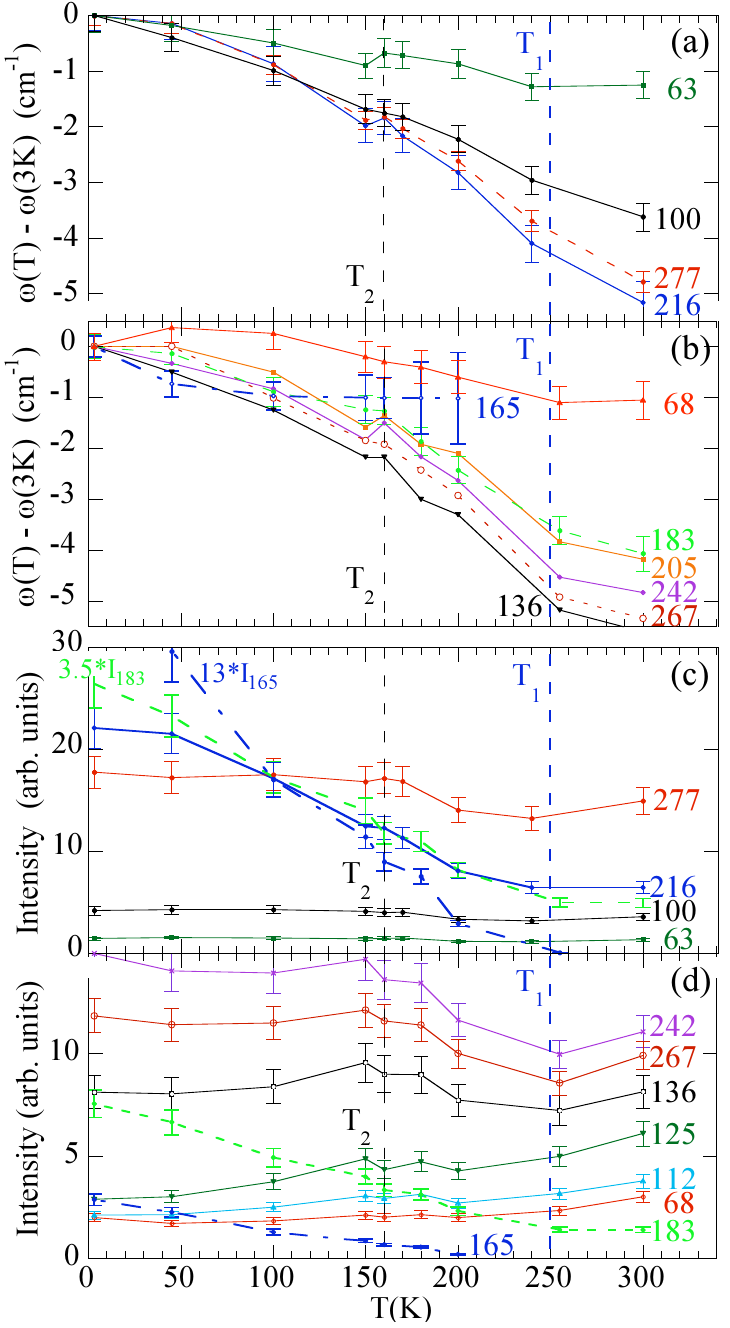}
\caption{(Color online) Temperature dependence of selected phonon frequencies (a,b) and intensities (c,d). Some error bars are omitted in (b) for clarity. $T_{1}$ and $T_{2}$ mark temperatures of structural phase transition at $\sim$ 250~K and phonon anomalies at $\sim$ 160~K. In (c) intensity of the Fe(2)-based B(B$_{g})\sim$ 216~cm$^{-1}$(solid blue line) scales almost perfectly to 3.5$\pm$~0.2 times intensity of Se(2)-based $A(A_{g})\sim$ 183~cm$^{-1}$ (dashed green line) in the range of 45 to 250~K.  It also satisfactory scales to 13$\pm$~1 times intensity of the Se(1)-based $A\sim$ 165~cm$^{-1}$(dot-dashed blue line) in the range of $\sim$ 100 to 200~K. Note there are four Se(2) per one Se(1) atom.}
\label{fig4}
\end{figure}

Temperature dependence of selected phonon frequencies and intensities are shown in Fig. \ref{fig4}(a,b) and \ref{fig4}(c,d), respectively. Apart from the structural phase transition at $T_{1}$ below $\sim$ 250~K, clearly there is a second characteristic temperature, $T_{2}\sim$ 160~K.
At $T_{2}$ majority of phonon vibration frequencies exhibit consistent discontinuity up to $\sim$ 0.3 cm$^{-1}$ [Fig. \ref{fig4}(a, b)], while quite a few modes display slop changes in their intensity vs temperature dependencies [Fig. \ref{fig4}(c, d)]. Since no new vibration modes (Raman or IR) are observed below the $T_{2}$ the $T_{2}$ is not constituent a structural phase transition, but rather referred as phonon anomaly temperature. Anomaly of a single $A_{g}$ mode at 66~cm$^{-1}$ at 160~K was mentioned by Zhang \textit{et al.} \cite{Raman1}. We would like to point out that the phonon anomalies seen at the $T_{2}\sim$ 160~K involve majority of measured Raman modes.

From experimental data at hand we could point out two implications of the observed structural phase transition on low-$T$ properties of K$_{0.75}$Fe$_{1.75}$Se$_{2}$. First, symmetry of the Se(1,2)-Fe(2) slab is broken at $T_{1}$, sample becomes ferroelectric, and $c$-axis structural distortions within the stab appears to build up on cooling. This is seen in Raman phonon peak intensities, Fig. \ref{fig4}(c,d), which are directly proportional to polarizability tensor. As sample enters the low-$T$ phase ($T<T_{1}$), polarizability of quite a few Raman-active modes build-up till $\sim T_{2}$, followed by saturation at $T<T_{2}$ ($B$-symmetry at 63, 100.6, and 277; $A$-symmetry at 81 and 267, Fano-shape 136 and 242~cm$^{-1}$) or reduction ($A$-symmetry at 68, 112, and 125~cm$^{-1}$). However, polarizabilities of Fe(2)-based $B$ mode at $\sim$ 216, As(2)-based $A$ mode at $\sim$ 183, and As(1)-based $A$ mode at $\sim$ 165~cm$^{-1}$ continue to build up till $\sim$ 45, 3, and 3~K, respectively. The scaling relationships among the three are shown in Fig. \ref{fig4}(c). The scaling is not surprising, giving to similar displacements patterns of the modes along the $c$-axis [Fig.\ref{fig1}(a1), (b1), and (b2)], so that their
Raman activities are driven by polarizability of the electronic orbitals forming the Fe-Se slabs. In iron-arsenides (122-systems) the Fe-As slab is perfectly symmetric and the Raman-active As-based $A_{1g}$ mode has extremely low intensity (polarizability) if measured in the same geometry \cite{modes122}.
It becomes visible upon doping destroying the symmetry of the slab. In K$_{0.75}$Fe$_{1.75}$Se$_{2}$, intrinsic population of Fe- and K- vacancies makes the Se(2)-based $A_{g}\sim$ 183~cm$^{-1}$ mode effortlessly visible at room temperature. The atomic displacements associated with
the \textit{I4/m} $\to $ \textit{I4} phase transition which have sizable $ab$-plane components are likely quenched below $T_{2}$, while $c$- axis displacements continue to build up on cooling. Second implication concerns the temperature dependence of low-frequency optical conductivity shown in Fig. \ref{fig1}(c). The conductivity displays weak temperature dependence above $T_{1}$ followed by faster increase below the $T_{2}$, in agreement with similar temperature dependence reported by Homes \textit{et al}. \cite{optics2}

Onset of superconductivity at $\sim$ 30 K has little effect on phonons. Using 3 and 45~K data points, the upper estimated phonon energy shifts are
-0.3 $\pm$ 0.4~cm$^{-1}$ (\textbar $\Delta \omega $\textbar /$\omega \sim$ 0.44{\%}) for 67.6~cm$^{-1}$
and $+$0.6 $\pm$ 0.4 cm$^{-1}$ (\textbar $\Delta \omega $\textbar /$\omega \sim$ 0.36{\%}) for 165.0~cm$^{-1}$ modes. Small frequencies renormalization implies either weak \textit{e-ph} interaction or that the phonons used in our analysis belongs to the AFM phase in the phase separated models \cite{TEM, XRD, optics, NQR, INS}: spectator AFM phonons would not feel onset of the superconductivity, unless via the proximity effect.

\section{CONCLUSIONS}

Raman scattering and optical conductivity were used to determine lattice vibration frequencies of superconducting crystal K$_{0.75}$Fe$_{1.75}$Se$_{2}$ in temperature range from 3 to 300~K. 19 Raman-active and 12 IR-active modes are observed at 3~K. The $\sim$ 136, 143, 242, and 277~cm$^{-1}$ Raman and $\sim$ 208~ cm$^{-1}$ IR mode exhibit Fano-like shape. The Raman Fano modes are due to the vibration coupling to AFM spin fluctuations, while the IR- mode is coupled to charge carriers in low-frequency part of optical conductivity. Raman phonon linewidths contain approximately equal contributions of two-phonon lattice anharmonicity on one hand and bare self-energy and broadening due to intrinsic defects on the other hand. The K$_{0.75}$Fe$_{1.75}$Se$_{2}$ undergoes \textit{I4/m} (space group {\#}87) $\to $ \textit{I4} (space group {\#}79) structural phase transition at $T_{1}$ below $\sim$ 250~K. Several modes which are not Raman- and IR-active in the measured geometry in \textit{I4/m} become visible in \textit{I4} phase including Raman modes at $\sim$ 165, 201, and 211~cm$^{-1}$ and IR-active modes at $\sim$ 99, 171, and 246~cm$^{-1}$. Weak phonon anomalies are also observed at at $T_{2} \sim$ 160~K. Symmetry of the Se(1,2)-Fe(2) slab is broken at $T_{1}$. $ab$-plane structural distortions are likely quenched below $T_{2}$, while $c$-axis structural distortions within the slab continues to build up on cooling down to 3~K.

\section{ACKNOWLEDGMENTS}
Research at Rutgers was supported by the U.S. DOE, office of BES, Division of Materials Science and Engineering under award DE-SC0005463. 
Research at Beijing National Laboratory for Condensed Matter Physics was supported by the NSFC and 973 projects of MOST 
(Grant No. 2011CB921701, 2012CB821403).

\end{document}